\documentclass[runningheads]{llncs}

\pdfoutput=1
\usepackage{times}
\usepackage{soul}
\usepackage{url}
\usepackage[hidelinks]{hyperref}
\usepackage[utf8]{inputenc}
\usepackage[small]{caption}
\usepackage{graphicx}
\usepackage{amsmath}
\usepackage{booktabs}
\usepackage{algorithm}
\usepackage{algorithmic}
\usepackage[switch]{lineno}

\usepackage{ulem}
\usepackage{mdframed}
\usepackage{multicol}
\usepackage{multirow}
\usepackage{amssymb}
\usepackage{threeparttable}
\usepackage{subfigure}
\usepackage{fancyhdr}
\usepackage{xcolor}
\usepackage{pifont}

\usepackage[misc]{ifsym} 

\usepackage{pgfplots}
\usepackage{colortbl}
\usepackage{color}

\usepackage{tikz}
\usetikzlibrary{arrows}
\usetikzlibrary{math}
\usetikzlibrary{positioning}

\usetikzlibrary{quotes,angles}
\usetikzlibrary{calc}
\usetikzlibrary{decorations.pathreplacing}
\usetikzlibrary{shapes.arrows}
\usepackage{verbatim}
\usepackage{pgfplots} 
\usepackage{pgfplotstable}
\pgfplotsset{compat=1.15}
\usetikzlibrary{patterns}

\definecolor{brandeisblue}{rgb}{0.0, 0.44, 1.0}

\def\ie{\textit{i.e.}} 

\def\eg{\textit{e.g.}}
\def\etal{\textit{et~al. }}

\newcommand{\sys}{CryptoMask}

\newcommand{\boolshare}[1] {\langle #1 \rangle}
\newcommand{\notejbai}[1]{{\color{orange}[\textsc{jbai:} #1]}}

\newcommand{\notescui}[1]{{\color{red}[\textsc{scui:} #1]}}

\begin{document}
\title{CryptoMask: Privacy-preserving Face Recognition}
%
%
\author{}
\institute{}

\author{Jianli Bai\inst{1} \and
Xiaowu Zhang\inst{2} \and
Xiangfu Song\inst{3,}\textsuperscript{(\Letter)} \and
Hang Shao\inst{4} \and
Qifan Wang\inst{1} \and
Shujie Cui\inst{5} \and
Giovanni Russello\inst{1} 
}
\authorrunning{J. Bai et al.}
%
\institute{University of Auckland, Auckland, New Zealand \\
\email{\{jbai795,qwan301\}@aucklanduni.ac.nz}~~ 
\email{g.russello@auckland.ac.nz}\\ \and
CloudWalk Technology, Beijing, China \\
\email{zhangxiaowu@cloudwalk.com} \\ \and
National University of Singapore, Singapore, Singapore \\
\email{songxf@comp.nus.edu.sg} \\ \and 
Beijing Institute of Graphic Communication, Beijing, China \\
\email{mir$\_$soh@163.com} \\ \and 
Monash University, Melbourne, Australia \\
\email{shujie.cui@monash.edu}}

%
\maketitle              
\setcounter{footnote}{0}
\begin{abstract}

Face recognition is a widely-used technique for identification or verification, where a verifier checks whether a face image matches anyone stored in a database. However, in scenarios where the database is held by a third party, such as a cloud server, both parties are concerned about data privacy. 
To address this concern, we propose \sys, a privacy-preserving face recognition system that employs homomorphic encryption (HE) and secure multi-party computation (MPC). We design a new encoding strategy that leverages HE properties to reduce communication costs and enable efficient similarity checks between face images, without expensive homomorphic rotation. Additionally, \sys~leaks less information than existing state-of-the-art approaches.
\sys~only reveals whether there is an image matching the query or not, whereas existing approaches additionally leak sensitive intermediate distance information. We conduct extensive experiments that demonstrate \sys's superior performance in terms of computation and communication. For a database with 100 million 512-dimensional face vectors, \sys~offers ${\thicksim}5 \times$ and ${\thicksim}144 \times$  speed-ups in terms of computation and communication, respectively.

\end{abstract}

\section{Introduction}\label{sec:introduction}

Biometric authentication has become increasingly vital in various applications in recent years. This work focuses on face recognition, which identifies or verifies a person's identity based on their facial features. Due to its ease of use and convenience, face recognition has gained significant traction in real-world applications such as public place surveillance (\eg, streets, airports, etc.)~\cite{parmar2014face}, social media~\cite{cherepanova2021lowkey}, and corporate punch card supervision~\cite{haigh2001chromium}.


As face recognition systems become more widespread, concerns about privacy have grown. In a typical system, a server stores face images belonging to users who are registered. When a verifier, who possesses a user's face image, queries the server to check if the user is verified, the system measures the similarity or distance between the queried image and the images in the database. However, in many cases, it may not be permissible to disclose users' face images to the server due to privacy concerns or the possibility of human rights abuses~\cite{bowyer2004face}. \textit{Therefore, it is essential to develop privacy-preserving face recognition protocols that protect data privacy while maintaining efficient recognition}.


Encrypting pre-processed images (\eg,~extracted face vectors) and performing face recognition over encrypted data is a straightforward approach to ensure data privacy. Homomorphic Encryption (HE) is a promising encryption scheme for this purpose, which was first proposed in \cite{rivest1978data} and realized in \cite{gentry2009fully}. HE allows computation in the encrypted domain without decryption. However, HE-based privacy-preserving face recognition protocols, such as the one proposed in \cite{boddeti2018secure}, are several orders of magnitudes slower than the original method, even when utilizing the Single-Instruction-Multiple-Data (SIMD) technique \cite{smart2014fully} to amortize the cost of homomorphic operations. To overcome this, the approach proposed in \cite{engelsma2022hers} explores encoding methods on the image database, reducing the number of homomorphic multiplications and rotations required and improving computation efficiency.
Moreover, previous works \cite{boddeti2018secure,engelsma2022hers} in this field fail to protect the private information of the database, as they allow the verifier to learn sensitive distance or similarity information and the number of face images close to the queried one.


In this paper, we propose Cryptomask, an efficient privacy-preserving face recognition protocol that only reveals a single bit of information to the verifier, indicating whether the queried face image is present in the database. 
We propose a novel encoding method to encrypt the database in a compact manner, resulting in improved performance. For distance computation, we use efficient matrix multiplication techniques that avoid expensive homomorphic rotations. Additionally, we ensure the privacy of distance calculations by designing a secure result-revealing protocol and optimizing its efficiency. 
CryptoMask outperforms existing distance-based privacy-preserving biometric schemes constructed via HE in terms of computation and storage overhead, and information leakage. Table~\ref{table::comparison} provides a comparison of different schemes, showing that our approach requires the least number of HE multiplications and additions and has minimal information leakage.
We implement CryptoMask and compare its performance with existing works~\cite{boddeti2018secure} and~\cite{engelsma2022hers}.
In the case of a database with 100 million face images, CryptoMask outperforms others up to~${\thicksim}5 \times$ and~${\thicksim}144~\times$ in computation and communication, respectively.  
\begin{table*}[!ht]
    \footnotesize  
	\centering 
 \caption{S{\upshape ummary of existing privacy-preserving face recognition protocols}.}\label{table::comparison}
	\begin{threeparttable}
	\begin{tabular}{c|c|c|c|c|c}
		\toprule
		\textbf{Protocol} & 
        \textbf{Multiplication} & \textbf{Addition} & \textbf{Rotation} & \textbf{Memory} &
        \textbf{Leakage}\\
        \hline
        Na{\"\i}ve & $md$ & $m(d-1)$ & 0 & $O(md\ell)$& $\boldsymbol{A},\boldsymbol{b},\boldsymbol{\mathsf{d}},r$\\
        \hline
        Hu \etal~\cite{hu2018outsourced} & $md^3$ & $md^2(d-1)$ & 0 & $O(md^2)$ & $\boldsymbol{\mathsf{d}},m$ \\
        \hline
        Pradel \etal~\cite{pradel2021privacy} & $md$ & $m(d-1)$ & 0 & $O(mdN)$ & $\boldsymbol{\mathsf{d}},m$ \\
        \hline
        Boddeti \etal~\cite{boddeti2018secure} & $m$ & $m$log$_2d$ & $m$log$_2d$ & $O(mN)$ & $\boldsymbol{\mathsf{d}},m$\\
        \hline
        HERS~\cite{engelsma2022hers} & $\lceil \frac{m}{N}\rceil d $ & $\lceil \frac{m}{N}\rceil (d-1) $ & 0 & $O(dN\lceil \frac{m}{N} \rceil)$  & $\boldsymbol{\mathsf{d}},m$ \\
        \hline
        Erkin \etal~\cite{erkin2009privacy} & $m(d+2)$ & $2m(d-1)$ & 0 & $O(mdN)$ & $m$ \\
        \hline
        \textbf{CryptoMask} & $\lceil \frac{m}{N-d}\rceil d $ & $\lceil \frac{m}{N-d}\rceil d $ & 0 & $O( dN \lceil\frac{m}{N-d}\rceil)$ & $m$ \\
		\bottomrule
	\end{tabular}
	\footnotesize{$\boldsymbol{A}:$ database containing face vectors; $\boldsymbol{b}:$ queried face vector; $m:$ database size; $d:$ dimension of each face vector; $N:$ HE plaintext polynomial degree; $l:$ length of each element in face vector;   $\boldsymbol{\mathsf{d}}:$ distance vector; $r:$ face recognition result. The notation $\lceil x \rceil$ denotes rounding up to the nearest integer of $x$. na{\"\i}ve represents the face recognition performed in plaintext. 
	}
    \end{threeparttable}
    \vspace{-3mm}
\end{table*}

\subsection{Related Work}
The early work given in~\cite{ross2010visual} relies on secret sharing to authenticate face recognition. However, it cannot ensure the privacy of face images. 
There are some similar works~\cite{rao2008fingerprint,uludag2005fuzzy,lee2008new} working for biometric authentication. Another line is employing pattern recognition to protect the queried database~\cite{patel2015cancelable,jin2017ranking}. However, this method also fails to ensure the security of the database and the queried face image. Some works~\cite{shashank2008private,upmanyu2009efficient} employ secure multi-party computation (MPC)~\cite{yao1986generate} to achieve the privacy-preserving goals, yet they are communication costly due to multiple interactions between the participants. Homomorphic encryption~\cite{rivest1978data} allows computations to be performed over encrypted data without first decrypting it. Many face recognition protocols~\cite{upmanyu2009efficient2,erkin2009privacy,troncoso2013fully,boddeti2018secure,engelsma2022hers} based on HE have been proposed. Unfortunately, they either result in heavy computation~\cite{erkin2009privacy,troncoso2013fully,boddeti2018secure} or cannot provide full secrecy~(\eg~leakage of distance similarity)~\cite{boddeti2018secure,engelsma2022hers}. We fill this gap by employing HE to perform distance computations and utilizing MPC to do a secure result-revealing process. Compared with the state-of-the-art~\cite{engelsma2022hers}, our work reduces both the computation and communication while maintaining the privacy of not only inputs and outputs but also intermediate data.

\section{Background}\label{sec:background}
In this section, we describe the face recognition algorithm and introduce the encoding method for a given matrix. Then we present some cryptographic primitives we use.

\subsection{Face Recognition}
In a face recognition system, each face image is represented by a feature vector, we say a face vector.
The extraction algorithm usually consists of face detection, alignment, normalization, and feature extraction, which is out of the scope of this work. 
We assume the face vector of each image is ready to use. 
In fact, the face vector extracted from the facial images of the same person could be slightly different. 
Thus, for face recognition, we should compare the similarity between two face vectors rather than check the equality. 
A simple method is to use either the Euclidean distance~\cite{danielsson1980euclidean} or the cosine similarity~\cite{singhal2001modern} to measure the similarity between two face vectors. 
In this paper, we employ cosine similarity. 
Specifically, given two vectors $\tilde{\boldsymbol{a}}=(\tilde{a}^0, ..., \tilde{a}^{d-1}) \in \mathbb{Z}^{d}$ and $\tilde{\boldsymbol{b}} =(\tilde{b}^0, ..., \tilde{b}^{d-1})\in \mathbb{Z}^{d}$, their cosine similarity is $d(\tilde{\boldsymbol{a}},\tilde{\boldsymbol{b}})  =\frac{\sum_{i=0}^{d-1} {{\tilde{a}}^i {\tilde{b}}^i}}{\sqrt{\sum_{i=0}^{d-1} {{(\tilde{a}^i)}^2}} \sqrt{\sum_{i=0}^{d-1} {(\tilde{b}^i)^2}}}$.
%
%
By setting $a^i=\frac{\tilde{a}^i}{\lVert \tilde{\boldsymbol{a}} \rVert}$ and $b^i=\frac{\tilde{b}^i}{\lVert \tilde{\boldsymbol{b}} \rVert}$, which are the normalization representations, we can convert it to $d(\tilde{\boldsymbol{a}},\tilde{\boldsymbol{b}}) 
=\sum_{i=0}^{d-1} {a}^i b^i$. 
By doing so, $d(\tilde{\boldsymbol{a}},\tilde{\boldsymbol{b}})$ can be considered as the inner product of vector $\boldsymbol{a}=(a^0$, ..., $a^{d-1})$ and $\boldsymbol{b}=(b^0, ..., b^{d-1})$. 
Note that $a^i$ and $b^i$ can be pre-computed offline. 
A larger value of $d(\tilde{\boldsymbol{a}},\tilde{\boldsymbol{b}})$ means higher similarity between $\tilde{\boldsymbol{a}}$ and $\tilde{\boldsymbol{b}}$, and if it is greater than a threshold value, we say $\tilde{\boldsymbol{a}}$ and $\tilde{\boldsymbol{b}}$ matches with each other, \ie, they represent the same person. In the following of this paper, all the face vectors are normalization representations. 



\subsection{Encoding Method} 
Given a set of encrypted face vectors, computing the cosine similarity one by one is time-consuming. A promising method is computing that in parallel. 
The encoding method from Cheetah~\cite{huang2022cheetah} achieves the best paralleling performance. In the following, we briefly describe the encoding method in Cheetah~\cite{huang2022cheetah}.


Given a matrix $\boldsymbol{\mathsf{A}}= \{\boldsymbol{a}_0, \boldsymbol{a}_1,...,\boldsymbol{a}_{\tilde{m}-1}\} \in \mathbb{Z}^{\tilde{m} \times d}$ with $\tilde{m}$ rows and $d$ columns, where $\boldsymbol{a}_i=(a_i^0, ..., a_i^{d-1})$ and $0 \leq i \leq \tilde{m}-1$, it can be represented into a polynomial as 
\begin{footnotesize}
\begin{align*}
\pi(\boldsymbol{\mathsf{A}}) &= a_0^{d-1}X^0+a_0^{d-2}X^1+\cdots+a_0^0X^{d-1}\\
&+a_1^{d-1}X^{d}+a_1^{d-2}X^{d+1}+\cdots + a_1^0X^{2d-1} + \\
&\cdots\\
&+a_{\tilde{m}-1}^{d-1}X^{(\tilde{m}-1)d}+a_{\tilde{m}-1}^{d-2}X^{(\tilde{m}-1)d+1}+\cdots+ a_{\tilde{m}-1}^0X^{\tilde{m}d-1}.
\end{align*}
\end{footnotesize}

\noindent Given another polynomial
$\pi(\boldsymbol{b}) = b^0X^0+b^1X^1+\cdots+b^{d-1}X^{d-1}$, we can get polynomial $\pi(\boldsymbol{\mathsf{d}})$ by computing  $\pi(\boldsymbol{\mathsf{d}}) \leftarrow \pi(\boldsymbol{\mathsf{A}})*\pi(\boldsymbol{b})$, where $*$ denotes polynomial multiplication. It is notable that the coefficient of degree $X^{(i+1)d-1}$, where $i \in [0,\tilde{m}-1]$ in polynomial $\pi(\boldsymbol{\mathsf{d}})$ forms the dot product result of the $i$-th row vector from $\boldsymbol{\mathsf{A}}$ and the vector $\boldsymbol{b}$. 
The correctness comes from the fact that the elements order of each vector in matrix $\boldsymbol{\mathsf{A}}$ is revised when it is encoded into a polynomial. 
We refer readers to Cheetah~\cite{huang2022cheetah} to see the detailed proof of correctness. 


\subsection{Homomorphic Encryption}
HE~\cite{acar2018survey} allows us to compute over encrypted data where the result is indeed the encrypted version of the operations on the plaintext. In this work, we use a lattice-based HE: ring learning with errors~(RLWE)-based HE called BFV~\cite{fan2012somewhat}. We briefly describe the construction of BFV scheme. See~\cite{fan2012somewhat} for a detailed formal description and security definition.



\noindent \textit{BFV Scheme}. 
The plaintext space of BFV scheme is taken from $R_t = \mathbb{Z}_t/(x^N+1)$ which represents polynomials with degree less than $N$ where $N$ is a power of 2, with the coefficients modulo $t$. Similarly, the ciphertext is defined in a ring $R_q$ with the coefficients modulo $q$. We use symbols $\boxplus$ and $\boxtimes$ to represent homomorphic addition and homomorphic multiplication, respectively. The BFV scheme consists of the following algorithms:
\begin{itemize}
    \item  $(pk,sk) \leftarrow$ KeyGen($1^\lambda$): On input the security parameter $\lambda$, it generates a pair of keys $(pk,sk)$.
    \item  $ct \leftarrow$ Encrypt($pk,\boldsymbol{m}$): On input the public key $pk$ and the plaintext $\boldsymbol{m}$, it outputs the ciphertext $ct$. 
    \item  $\boldsymbol{m} \leftarrow$ Decrypt($sk,ct$): On input the secret key $sk$ and the ciphertext $ct$, it outputs a plaintext $\boldsymbol{m}$.
    \item Eval($ct_i,ct_j$): Given two ciphertexts $ct_i$ and $ct_j$, output a ciphertext corresponding to the following operation.
        \begin{itemize}
        \item[-] Eval.Add($ct_i$,$ct_j$): Output $ct \leftarrow ct_i \boxplus ct_j$. 
        \item[-] Eval.Mul($ct_i$,$ct_j$): Output $ct \leftarrow ct_i \boxtimes ct_j$. 
        \end{itemize}
\end{itemize}

\subsection{Key-switching}
Key-switching enables the data encrypted by one set of encryption keys to be re-encrypted by another without decrypting the data. BFV scheme~\cite{fan2012somewhat} naturally supports the key-switching operation.
The key-switching process consists of two algorithms:
\begin{itemize}
    \item $k_{A\rightarrow B} \leftarrow $ SwKeyGen($sk_A, sk_B$): On input two BFV secret keys $sk_A, sk_B$, it outputs a key-switching key $k_{A\rightarrow B}$.
    \item $ct_B \leftarrow$ Switching($ct_A,k_{A\rightarrow B}$): On input a key-switching key $k_{A\rightarrow B}$ and a ciphertext $ct_A$ encrypted by a public key $pk_A$ associated with $sk_A$, it outputs a ciphertext $ct_B$ encrypted by a public key $pk_B$ associated with $sk_B$.
\end{itemize}
More details about the key-switching technique can be found in~\cite{kim2021revisiting}.


\subsection{Secret Sharing} 
For an $l$-bit value $x \in \mathbb{Z}_{2^l}$, we use $\boolshare{x}^A$ to denote $x$ is arithmetically shared between parties $P_0$ and $P_1$ where $P_0$ holds $x_0^A$ and $P_1$ holds $x_1^A$ such that $x = x_0^A + x_1^A$ with $x_0^A$, $x_0^A \in \mathbb{Z}_{2^l}$. Similarly, $\boolshare{x}^B$ denotes a boolean share of $x$ where $x = x_0^B \oplus x_1^B$ with $x_0^B$, $x_0^B \in \mathbb{Z}_{2^l}$. Note that each share itself does not reveal any information about $x$. In some cases, we need the conversion between different sharing formats. 
We use the $\textbf{B2A}$ technique to convert $x$ from its boolean sharing $\boolshare{x}^B$ to its arithmetic sharing $\boolshare{x}^A$, which we represent as $(x_0^A, x_1^A) \leftarrow$ B2A$(x_0^B,x_1^B)$.
The detailed $\textbf{B2A}$ conversion can be referred to~\cite{demmler2015aby}. 
If $\boldsymbol{x}$ is a vector, then $\boldsymbol{x} = \boldsymbol{x}^A_0 + \boldsymbol{x}^A_1$ means each element in the vector is additionally shared between two parties. In our design, the cloud server (CS) plays the role of $P_0$, and the verifier plays the role of $P_1$.

\subsection{Secure Comparison} 
Secure comparison, also known as Millionaire's problem~\cite{yao1986generate}, compares two integers held by two parties. The inputs contain $x$ from one party and $y$ from another party, and the output bit 1 or 0 is shared between the two parties. Cryptflow2~\cite{rathee2020cryptflow2} proposes an efficient comparison protocol based on the observation: assume $x=x_1||x_0$ and $y=y_1||y_0$, we must have $x<y$ either when $x_1=y_1$ and $x_0<y_0$ or when $x_1<y_1$, \ie, $\bold{1}\{x<y\}=(\bold{1}\{x_1=y_1\}\wedge \bold{1}\{x_0<y_0\}) \oplus \bold{1}\{x_1<y_1\}$\footnote{$\bold{1}\{condition\}$ and $\bold{0}\{condition\}$ mean the condition is true and false, respectively.}. By separating the binary represented values into small parts, the queried Oblivious Transfer (OT)~\cite{ishai2003extending} is also small, optimizing the communication cost.
Recently, Cheetah~\cite{huang2022cheetah} provides further optimization by replacing the underlying secure AND implementation with Random OT (ROT)~\cite{ishai2003extending} generated Beaver Triples~\cite{beaver1992efficient}. For simplicity, we represent secure comparison as $(b_0,b_1)\leftarrow$ SC$_{lt}(x,y)$ which means one party inputs $x$ and another party inputs $y$ and outputs $b=1$ if $x<y$ and $b=0$ otherwise, where $b = b_0 \oplus b_1$. For more details about the state-of-the-art secure comparison, please refer to ~\cite{huang2022cheetah,rathee2020cryptflow2}.

\section{Overview of Our Approach} \label{sec:overview}
This section describes the system model and threat model and overviews CryptoMask.

\subsection{System Model}
\begin{figure}[tb]
	\centering 
    \includegraphics[height=1.8in]{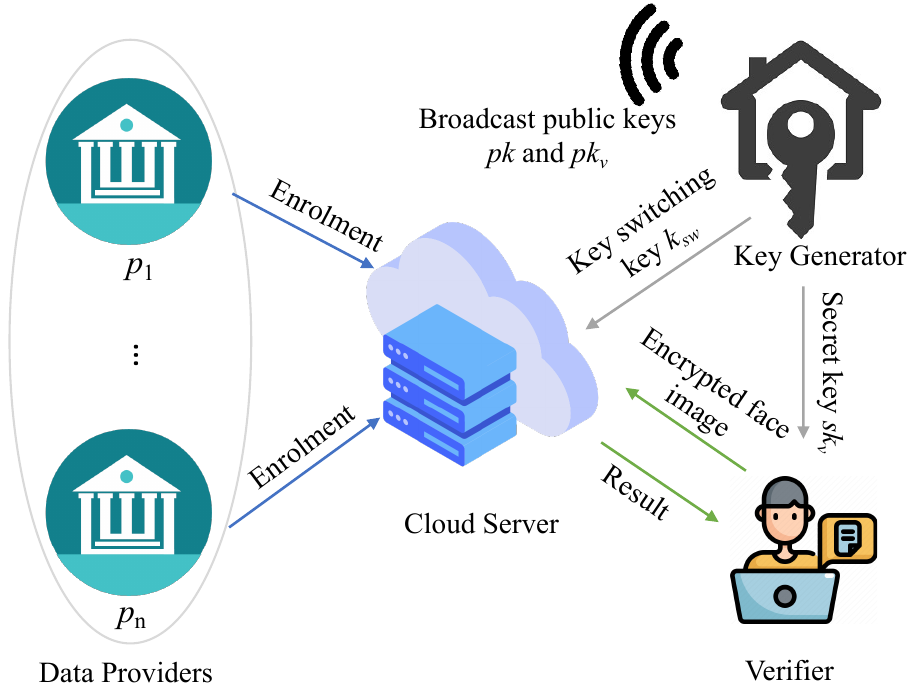} 
	\caption{System model.} 
	\label{fig:system} 
\end{figure}

In CryptoMask, we consider the scenario where the database is stored on a cloud server, and the corresponding face vectors are received from a group of data providers. 
A verifier wants to check if a given face image matches an image in the database.
Our system consists of four types of entities: a trusted \textbf{Key Generator~(KG)} who generates keys for other entities for privacy-preserving purposes. A group of \textbf{Data Providers~(DPs)} who upload extracted face vectors to a cloud server, a \textbf{Cloud Server (CS)} who stores the database of face vectors,
and a \textbf{Verifier} who checks if a given face vector is in the database, as shown in Fig.~\ref{fig:system}.

\noindent \textbf{KG.} KG generates a pair of HE public/private keys $(pk,sk)$ and distributes $pk$ to other entities. KG also generates another pair of public/private keys $(pk_v,sk_v)$ and sends them to the verifier. When KG receives a “setup" request from the verifier, it computes a key-switching key $k_{sw}$ based on $sk$ and $sk_v$ and sends it to CS.

\noindent \textbf{DPs.} In our system, DPs can upload images (represented by face vectors) to CS. To keep their data private, DPs encrypt the face vectors using the public key $pk$ before uploading them to CS. We call this process \textit{enrolment}. 

\noindent \textbf{CS.} CS stores the encrypted face vectors. 
It performs face recognition protocol with the verifier without learning anything about the queried face information or the result. 

\noindent \textbf{Verifier.} 
The verifier has a face image and intends to check if the image is in the database by performing a privacy-preserving face recognition protocol with CS. We call this process \textit{evaluation}.
It learns the image exists in the database if the check result is one.
For example, a verifier can be a service provider who receives or collects a face image from a user after the user's consent. The verifier then wants to check whether the user is a verified user in order to provide subsequent service. 

\subsection{Threat Model}
Similar to previous work, such as \cite{boddeti2018secure} and~\cite{engelsma2022hers}, we assume the CS and the verifier are honest-but-curious~(semi-honest). 
That is, they will follow the protocol honestly but may try to infer as much information as possible. We also assume CS and the verifier will never collude with each other. It is reasonable in practice because CS (\eg, education management organization) is motivated to maintain its reputation and is not likely to take the risk of colluding with the verifier. The KG is a fully trusted party.

\subsection{Overview of \sys}

Encrypting each face vector with HE and computing the cosine similarity between the query and each vector in the database is a straightforward but expensive way to perform face recognition securely. With $m$ face vectors and $d$ features per vector, this method requires $md$ homomorphic multiplications, which can be significantly time-consuming. Additionally, this approach poses a privacy risk by leaking sensitive information, such as the computed distance vectors $\boldsymbol{\mathsf{d}}$. 
Previous works, such as those proposed in~\cite{boddeti2018secure,engelsma2022hers,erkin2009privacy,pradel2021privacy}, also suffer from the same issue. 
To tackle all the issues above, we introduce \sys. In particular, we design a novel encoding method to enhance performance and a secure result-revealing protocol to minimize information leakage.

To reduce both the communication and computation overhead, our main idea is to encrypt face vectors in batches and compute the cosine similarity between the query and a batch of face vectors, rather than one by one. 
Specifically, during the enrollment process, given a batch of face vectors, DP encodes them into one BFV ciphertext $ct_i$ and sends it to CS. 
When the verifier queries for an image, CS performs only one homomorphic multiplication between each BFV ciphertext $ct_i$ and the encrypted query. The resulting ciphertext contains the cosine similarity between batched face vectors and the queried face vector. 
To determine if the queried image matches any image stored in the CS, the next step is to compare the cosine similarity with the threshold. 
Directly revealing the cosine similarity results to the verifier or the CS exposes sensitive information. For example, they can learn how many face images in the database are similar to the given one. 
To avoid such leakage, CryptoMask runs a secure result-revealing protocol between CS and the verifier, which only reveals whether the queried face image exists in the database to the verifier. 

To further enhance the performance of \sys, we can adopt a paralleling technique to compute the cosine similarity between the query and batched face vectors.
As done in work~\cite{boddeti2018secure}~\cite{engelsma2022hers}, the homomorphic multiplication performed during the evaluation can be processed in parallel with the SIMD technique. 
However, this technique requires a prime plaintext modulus\cite{huang2022cheetah}, implying that the homomorphic encryption must be performed in $\mathbb{Z}_p$ with $p$ as prime. In our secure result-revealing protocol, the secure comparison is a non-linear function, and~\cite{rathee2020cryptflow2} has shown that OT-based protocols on the ring $\mathbb{Z}_{2^l}$ perform 40\%-60\% better than on the prime field $\mathbb{Z}_p$ in bandwidth consumption, with almost no cost for modulo reduction.
Hence, in this work, rather than employing SIMD, we opt for the parallelization technique from~\cite{huang2022cheetah} to compute homomorphic multiplication in parallel. This technique enables us to work exclusively in the ring domain $\mathbb{Z}_{2^l}$ and brings another efficiency improvement by avoiding expensive rotation, the key operation for SMID-based work.
Furthermore, while~\cite{huang2022cheetah} necessitates an extraction algorithm (RLWE-based ciphertext to LWE-based ciphertext) for useful information extraction from the resulting ciphertext, we avoid it by masking the resulting ciphertext and sending it back to the verifier, which is more efficient.

\subsection{Data Representation}
The coefficients of the BFV plaintext polynomial must be integers. To achieve this, we need to encode our real-valued representation $\boldsymbol{\mathsf{A}} \in \mathbb{R}^{m\times d}$ as an integer-valued representation, which we denote by $\boldsymbol{\mathsf{A}} \in \mathbb{Z}^{m\times d}$. For the remainder of the paper, we use $\boldsymbol{\mathsf{A}}$ to refer to the matrix where all elements are integers. We scale the real-valued features into integers using a specified precision. This scaling method results in a loss of precision during computation. In our experiments, we evaluate the level of precision loss by setting different precision scales, and report the results in Table~\ref{Table::Accuracy} in Appendix~\ref{appendix:accuracy}.

\section{CryptoMask Details} \label{sec:approach}
This section describes the enrollment and evaluation processes of CryptoMask in detail. 


\subsection{Our Encoding Method}
BFV scheme~\cite{fan2012somewhat} is designed to work on a polynomial ring $R_t = \mathbb{Z}_t/(x^N+1)$ with degree $N$. 
The observation is that the number of slots in a polynomial (\eg, 4096) is far more than the dimension of a face vector (\eg, $d=128$). Thus, we can employ one polynomial to represent multiple face vectors as done in Cheetah~\cite{huang2022cheetah}. 
In our design, each row in the matrix $\boldsymbol{\mathsf{A}}$ represents a face vector. That is, before encrypting and uploading the face vectors to CS, DP encodes them into a matrix $\boldsymbol{\mathsf{A}}$ and then transforms it into the polynomial $\pi(\boldsymbol{\mathsf{A}})$. 
Then DP encrypts this polynomial using BFV as $ct_i$ and sends it to CS. The verifier encrypts the queried face vector $\boldsymbol{b}$ as $ct$ and sends it to CS. The cosine similarity is computed by multiplying these two ciphertexts $ct_i$ and $ct$, whose underlying plaintext polynomial is exactly $\pi(\boldsymbol{\mathsf{d}})$.
As mentioned, the plaintext space of BFV scheme is taken from $R_t = \mathbb{Z}_t/(x^N+1)$, which means the maximum degree of a plaintext polynomial is $N$. The direct method is we fill all the coefficients slots in the plaintext polynomial when considering encoding our face vectors database. However, this might result in a loss of valid similarity.
The reason is that the valid value in the product will be dropped~(module reduced to a position with a degree less than $N$) if its associated degree is greater than $N$, which means we will get the wrong distance between the last face vector in the matrix and the queried image. Our idea is to leave the last $d$ positions in the polynomial $\pi(\boldsymbol{\mathsf{A}})$ for “buffer” use and set their coefficients as 0. Thus, all valid values will be presented as coefficients with degrees less than $N$.
That is, if the degree of a plaintext polynomial is $N$, we only encode its lower $N-d$ coefficients and leave the higher $d$ coefficients as zeros. A similar strategy applies to the queried face vector.
Using this encoding method, the concrete number of ciphertext for $m$ face vectors with dimension $d$ will be $\lceil \frac{md^2}{N-d} \rceil$.

\begin{algorithm}[tb]
    \footnotesize
    \caption{Secure enrolment}
    \label{alg:enrolment}
    \textbf{Input}: An indicator $ind$ and the last ciphertext $ct_{la}$ from CS; $n_u$ $d$-dimensional face vectors $\boldsymbol{\mathsf{V}} = \{\boldsymbol{a}_0,\cdots, \boldsymbol{a}_{n_u-1}\} \in \mathbb{Z}^{n_u\times d}$ and public key $pk$ from DP.\\
    \textbf{Parameter}: $\delta =  \lceil \frac{N-d}{d} \rceil$ where $N$ is the plaintext polynomial degree.\\
    \textbf{Output}: CS adds the encrypted face vectors to the database.
    
    \begin{algorithmic}[1] 
        \STATE DP informs CS to add new face vectors. CS sends $ind$ to DP.
		
		\STATE DP takes $\delta-ind$ face vectors and organizes them into a matrix $\boldsymbol{\mathsf{A}}_0 \in \mathbb{Z}^{\delta \times d}$ by padding $ind$ zero vectors before these real samples. Then DP represents $\boldsymbol{\mathsf{A}}_0$ as $\pi(\boldsymbol{\mathsf{A}}_0)$ and gets $ct_0 \leftarrow$  Encrypt$(pk,\pi(\boldsymbol{\mathsf{A}}_0))$.
		
		\STATE DP separates the remaining vectors into $e\delta$ vectors and remains $f$ vectors where $f < \delta$ and $n_u = \delta -ind + e \delta + f$. 
        \STATE DP constructs $e$ polynomials $\pi(\boldsymbol{\mathsf{A}}_1), \cdots, \pi(\boldsymbol{\mathsf{A}}_e)$ using $e\delta$ face vectors and performs $ct_i \leftarrow$ Encrypt$ (pk,\pi(\boldsymbol{\mathsf{A}}_i))$ for each $i \in [1,e]$.
		
		\STATE DP pads $\delta -f$ zero vectors to the remaining $f$ vectors and gets $\pi(\boldsymbol{\mathsf{A}}_{e+1})$. Then DP encrypts it as $ct_{e+1} \leftarrow$ Encrypt$(pk,\pi(\boldsymbol{\mathsf{A}}_{e+1}))$ and sets $ind \leftarrow \delta -f$.

		\STATE DP uploads $\{ct_0,\cdots,ct_{e+1} \}$ and $ind$ to CS.
		
		\STATE After receiving the ciphertexts, CS first updates $ind$ and saves $\{ct_1,\cdots,ct_{e+1} \}$. Then CS performs $ct_{la} \leftarrow$ Eval.Add($ct_{la},ct_0$). 
    \end{algorithmic}
\end{algorithm}

\begin{algorithm}[tb]
\footnotesize
    \caption{Secure distance computation}
    \label{alg:distance computation}
    \textbf{Input}: An encrypted database $\{ct_0,\cdots,ct_{s-1}\}$, where each $ct_i$ is the ciphertext of a $\delta \times d$ matrix $\boldsymbol{\mathsf{A}} = \{\boldsymbol{a}_0, \boldsymbol{a}_1,\cdots,\boldsymbol{a}_{\delta-1}\} \in \mathbb{Z}^{\delta \times d}$ with $m=s\delta$; A queried face vector $\boldsymbol{b} \in \mathbb{Z}^d$ from verifier.\\
    \textbf{Parameter}: $\delta =  \lceil \frac{N-d}{d} \rceil$ where $N$ is the plaintext polynomial degree.\\
    \textbf{Output}: CS gets the secret share $\boldsymbol{\mathbf{d}}^A_0$ and the verifier gets the secret share $\boldsymbol{\mathbf{d}}^A_1$ where $\boldsymbol{\mathbf{d}} = \boldsymbol{\mathbf{d}}^A_0 + \boldsymbol{\mathbf{d}}^A_1$ is an $m$-length vector of computed distances.
    
    \begin{algorithmic}[1] 

        \STATE The verifier sends a ``setup" signal to KG. Then KG generates a key-switching key $k_{sw} \leftarrow $ SwKeyGen($sk, sk_v$) and sends it to CS.
    
        \STATE The verifier encodes and encrypts $\boldsymbol{b}$ as $ct \leftarrow$ Encrypt$(pk,\pi(\boldsymbol{b}))$ and sends $ct$ to CS.

        \STATE CS and the verifier generate two empty vectors $\boldsymbol{\mathbf{d}}^A_0$ and $\boldsymbol{\mathbf{d}}^A_1$, respectively.
		
		\FOR{$i \in [0,s-1]$}
		    \STATE CS computes $ct_i' \leftarrow$ Eval.Mul($ct$, $ct_i$). 
		    
		    \STATE CS randomly generates a plaintext polynomial $\boldsymbol{r}_i = r_0X^0 + \cdots+ r_{N-1}X^{N-1}$. 
		    
		    \STATE CS extracts its the $(kd-1)$-th coefficients from $\boldsymbol{r}_i$ and sets $\boldsymbol{\mathbf{d}}^A_0 [i\delta+k-1] \leftarrow -r_{kd-1}$ where $k \in [1,\delta]$.
		    
		    \STATE CS computes $ct_i'' \leftarrow$ Eval.Add($\boldsymbol{r}_i$, $ct_i'$).

            \STATE CS performs $c_i' \leftarrow$Switching$(k_{sw},ct_i'')$.
            
		\ENDFOR

		\STATE CS sends $\{c_0',\cdots, c_{s-1}'\}$ to the verifier and keeps $\boldsymbol{\mathbf{d}}^A_0 [i]$ where $i \in [0,m-1]$.

		
		\FOR{$i \in [0,s-1]$}

            \STATE The verifier performs $\boldsymbol{p}_i \leftarrow$ Decrypt$(sk_v, ct_i')$ for each $i \in[0,s-1]$, where $\boldsymbol{p}_i = a_0X^0 + \cdots+ a_{N-1}X^{N-1}$.
            
		    \STATE The verifier extracts the $(kd-1)$-th coefficients $a_{(kd-1)}$ from polynomial $\boldsymbol{p}_i$ and sets $\boldsymbol{\mathbf{d}}^A_1 [i\delta+k-1] \leftarrow a_{kd-1}$ where $k \in [1,\delta]$.
        \ENDFOR
        
    \end{algorithmic}
\end{algorithm}

\begin{algorithm}[tb]
\footnotesize
    \caption{Secure result-revealing}
    \label{alg:result-revealing}
    \textbf{Input}: The secret share of distance vector $\boldsymbol{\mathbf{d}}^A_0$ from CS; The secret share of distance vector $\boldsymbol{\mathbf{d}}^A_1$ and a threshold $ts$ from verifier.\\
    \textbf{Output}: The verifier learns whether its face image exists in the database.
    
    \begin{algorithmic}[1] 
        
        \FOR{$i \in [0,m-1]$}

        \STATE The verifier updates $\boldsymbol{\mathbf{d}}^A_1 [i] \leftarrow ts-\boldsymbol{\mathbf{d}}^A_1 [i]$.
        
        \STATE CS and verifier jointly run $(b^B_0[i],b^B_1[i]) \leftarrow$ SC$_{lt} (\boldsymbol{\mathbf{d}}^A_1 [i],\boldsymbol{\mathbf{d}}^A_0 [i])$.

        \STATE CS and verifier jointly perform $(b^A_0[i], b^A_1[i]) \leftarrow$ B2A$(b^B_0[i],b^B_1[i])$. 


        \ENDFOR
        
        \STATE CS computes $b_0 = \sum^{m-1}_{i=0}b^A_0[i]$ and verifier computes $b_1 = \sum^{m-1}_{i=0}b^A_1[i]$. 
        
        \STATE CS and verifier jointly perform $(\mu_0,\mu_1) \leftarrow$ SC$_{lt} (-b_0, b_1)$.

        \STATE CS sends $\mu_0$ to the verifier. The verifier computes $\mu \leftarrow \mu_0 \oplus \mu_1$ and learns its face image is in the database by $\mu=1$. Otherwise, it learns its face image is not in the database by $\mu=0$.
    \end{algorithmic}
\end{algorithm}

\subsection{Enrolment Process}

Based on our encoding method, we improve enrollment efficiency by reducing the number of ciphertexts uploaded by DPs. Specifically, we use one plaintext polynomial with degree $N$ to represent $\lceil \frac{N-d}{d} \rceil$ face vectors, which results in only one homomorphic ciphertext. 
Thus, DP only needs to upload a single homomorphic ciphertext for $\lceil \frac{N-d}{d} \rceil$ images to CS while the state-of-the-arts~\cite{boddeti2018secure} and~\cite{engelsma2022hers} require $\lceil \frac{N-d}{d} \rceil$ and $d$ ciphertext, respectively. This encoding strategy is also beneficial to CS for saving storage overhead compared with work~\cite{boddeti2018secure}, \cite{engelsma2022hers}. The reason is that our designed encoding method allows CS to merge its last stored ciphertext with a new one that comes from another DP.

The details of the enrollment process are given in Algorithm~\ref{alg:enrolment}. We suppose CS already stored some encrypted face vectors under the public key $pk$ and a DP then wants to add $n_u$ $d$-dimensional face vectors $\boldsymbol{\mathsf{V}} = \{\boldsymbol{a}_0,\cdots,\boldsymbol{a}_{n_u-1}\}$ to CS. CS maintains an indicator $ind$, which tells DP the start vacant position in the last stored ciphertext. Rather than directly encrypting these vectors and sending them to CS, DP first encodes the data based on our proposed encoding method and then performs BFV encryption over the encoded data. When receiving the indicator $ind$ from CS, DP divides its vectors into three parts. The first part contains $\delta-ind$ vectors where $\delta =  \lceil \frac{N-d}{d} \rceil$ represents the maximum number of face vectors that can be encoded into a polynomial. Since CS is allowed to merge the last ciphertext with a newly come one, DP organizes the first $\delta-ind$ vectors into a matrix $\boldsymbol{\mathsf{A}}_0 \in \mathbb{Z}^{\delta \times d}$ by padding $ind$ zero vectors before these real samples. This matrix is encrypted as $ct_0$. When CS receives $ct_0$, it can merge it to its last stored ciphertext $ct_{la}$ by simply performing a homomorphic addition Eval.Add($ct_{la},ct_0$). The second part contains $e \delta$ vectors, and the last part contains $f$ vectors where $f < \delta$ and $n_u = \delta -ind + e \delta + f$. DP encrypts matrices $\boldsymbol{\mathsf{A}}_1, \cdots, \boldsymbol{\mathsf{A}}_e$ in the second part separately using BFV and sends them to CS. Unlike the first part, for the last part, DP first pads $\delta -f$ zero vectors to the remaining vectors, then encrypts it as $ct_{e+1}$ and sends it to CS. In the last, DP updates the indicator $ind = \delta -f$ and sends it to CS for further use. 

\subsection{Evaluation Process}
The evaluation process happens between a verifier and a CS. Specifically, as shown in Algorithm~\ref{alg:distance computation}, given a face vector, the verifier first encodes it into a polynomial. Then the verifier encrypts the polynomial using the public key $pk$ and sends it to CS. CS gets a key-switching key $k_{sw}$ from KG after KG receives a ``setup" signal from the verifier. After receiving the encrypted query $ct$ from the verifier, CS first runs local homomorphic multiplication between each ciphertext $ct_i$ stored in CS and $ct$, where $i \in [0,s-1]$. Rather than directly sending the computed results to the verifier, CS masks each of them using a randomly selected plaintext polynomial $\boldsymbol{r}_i$. CS can easily extract the $(kd-1)$-th coefficients $r_{kd-1}$ from $\boldsymbol{r}_i$ where $k \in [1, \delta]$ and keeps its additive inverse into $\boldsymbol{\mathbf{d}}^A_0[i\delta+k-1]$, which is one of the secret parts of computed distances. To enable the verifier to perform decryption by itself, CS transfers each ciphertext encrypted by $pk$ to $pk_v$ by a key-switching technique before sending them to the verifier. With the masked distances, CS does not require performing RLWE to LWE extraction function, a key design in~\cite{huang2022cheetah}. The extraction function is considered time-consuming as it is performed over homomorphic ciphertext~\cite{chen2021efficient}. In our design, the verifier can extract the coefficients by itself after decrypting the RLWE ciphertext. Doing this saves the homomorphic extraction overhead on the CS side. Besides, we also reduce the required communication for $\lceil \frac{N-d}{d}\rceil$ face vectors from $\lceil \frac{N-d}{d}\rceil(N+1)q$ to $2Nq$ where $q$ denotes the ciphertext coefficients modulo. After decrypting all the received ciphertext, the verifier similarly extracts coefficients from obtained polynomial and saves them into $\boldsymbol{\mathbf{d}}^A_1$, which is another part of secret-shared computed distances.

Then CS runs a secure result-revealing protocol with the verifier as shown in Algorithm~\ref{alg:result-revealing}. For each shared distance, CS and the verifier jointly run a secure comparison to compute $\boldsymbol{\mathbf{d}}^A_1 [i] < \boldsymbol{\mathbf{d}}^A_0 [i]$, where $\boldsymbol{\mathbf{d}}^A_1 [i] \leftarrow ts - \boldsymbol{\mathbf{d}}^A_1 [i]$ is from the verifier and $\boldsymbol{\mathbf{d}}^A_0 [i]$ is from CS. Clearly, the result represents the less than comparison between the given threshold $ts$ and the distance $\boldsymbol{\mathbf{d}}[i]$. However, the comparison result is in binary format, so we cannot directly aggregate all results. Thus, we need a \textbf{B2A} conversion $(b^A_0[i], b^A_1[i]) \leftarrow$ B2A$(b^B_0[i],b^B_1[i])$. After that, CS can compute $b_0 = \sum^{m-1}_{i=0}b^A_0[i]$ and the verifier computes $b_1 = \sum^{m-1}_{i=0}b^A_1[i]$. To obtain the queried result, CS and verifier jointly perform $(\mu_0,\mu_1) \leftarrow$ SC$_{lt} (-b_0, b_1)$ and CS sends $\mu_0$ to the verifier. In the end, the verifier learns whether the queried face image exists in the database by computing $\mu \leftarrow \mu_0 \oplus \mu_1$.  

\subsection{Security Analysis}
The security of CryptoMask follows from the semantic security of HE and the security of MPC. The complexity and security analysis can be found in Appendix~\ref{sec:security}.

\subsection{Optimizations}\label{subsec::optimization}
We present some optimizations to improve the efficiency of CryptoMask. 

\noindent \textbf{Reducing Computation Overhead.}
In Algorithm~\ref{alg:distance computation}, CS should run a key-switching before sending the masked distance ciphertext to the verifier, which is time-consuming. We can put this key-switching when the verifier first sets up. Rather than sending the face vector encrypted by $pk$, the verifier encrypts it using its public key $pk_v$. Then all computations in CS are over the encrypted data over $pk_v$. However, this is a trade-off since it will save computation overhead but increase CS's storage.

\noindent \textbf{Reducing Communication Overhead.}
We employ the ciphertext compression technique from SEAL library~\cite{sealcrypto}, compressing the original ciphertext into around two-thirds of the original size. Notably, this ciphertext compression can only be used for data to be decrypted because it will cause a decryption error if the data is computed over compressed ciphertext. Clearly, CryptoMask can benefit from the compression technique.
Another ciphertext size reduction of CryptoMask is gained from Cheetah~\cite{huang2022cheetah}. The observation is that CS only needs to send high-end bits of two parts of ciphertext to the verifier. In this way, we save around $16\%-25\%$ communication with a negligible decryption failing chance~(\ie, $< 2^{-38}$). For a more detailed analysis, see~\cite{huang2022cheetah}. 


\section{Performance Evaluation}
\label{sec:experiment}
We implemented a prototype of \sys~on top of Cheetah~\cite{huang2022cheetah} and evaluated its performance with different datasets. 
In this section, we present our experimental results.

\noindent \textbf{Experimental Setup.} 
The experiment runs on a laptop running Centos 7.9 equipped with Xeon(R) Gold 6240 2.6GHZ CPU with 32~GB RAM. 
The network setting is LAN with RTT 0.1~ms and bandwidth 1~Gbps.
We run all the experiments in a single-threaded environment.
We set the BFV parameter $N$ as 4096, $t$ as 20 bits, and $q$ as $60+49$ bits. The security level $\lambda$ is set as 128 bits. 
We also evaluated the performance of the existing works~\cite{boddeti2018secure} and \cite{engelsma2022hers} in the same environment with the same values for parameters. We compared their results with \sys. The time we report is averaged over ten trials. 


\noindent \textbf{Datasets.} 
Similar to~\cite{boddeti2018secure} and~\cite{engelsma2022hers}, we evaluate the performance of CryptoMask with datasets that have different numbers of face images and dimensions. 
To show how the accuracy is influenced by precision scaling, as done in~\cite{boddeti2018secure}, we use a real dataset LFW~\cite{huang2008labeled} for the evaluation, which can be obtained from~\cite{LFWTech}. 
Specifically, LFW consists of 13,233 face images of 5,749 subjects. 
As done in~\cite{boddeti2018secure} and~\cite{engelsma2022hers}, We utilize the state-of-the-art face representation FaceNet~\cite{schroff2015facenet} to extract face vectors.



\begin{figure*}[tb]
\centering
    \subfigure[32-D Representation]{
        \begin{tikzpicture}[scale=0.5]
        \begin{axis}[%
        ylabel=Number,
        scaled ticks=false, 
        legend pos= north west, 
        grid=major,
        grid style=dashed,
        xmin= 1,
        xmax= 10^6,
        ymin= 0,
        ymax= 100000,
        xlabel={Database Size},
        xmode=log,
        log basis x={10},
        ymode=log,
        log basis y={10},
        ylabel={Running Time (s)},
        ]

        \addplot[color=green,mark=square] coordinates {
                (1, 0.02452)
                (10, 0.21295)
                (100, 2.04535)
                (1000, 20.3486)
                (10000, 204.531)
                (100000, 2012.381)
                (1000000, 21298.926)
            };
        \addlegendentry{SFM}
        
        \addplot[color=blue,mark=square] coordinates {
                (1, 0.225089)
                (10, 0.224039)
                (100, 0.219675)
                (1000, 0.225891)
                (10000, 0.678267) 
                (100000, 5.512802)
                (1000000, 55.579168)
            };
        \addlegendentry{HERS}

        \addplot[color=red,mark=square] coordinates {
                (1, 0.012)
                (10, 0.02)
                (100, 0.025)
                (1000, 0.084)
                (10000, 0.562)
                (100000, 5.861)
                (1000000, 75.048)
            };
        \addlegendentry{CryptoMask-W}

        \addplot[color=brandeisblue,mark=square] coordinates {
                (1, 0.008)
                (10, 0.008)
                (100, 0.007)
                (1000, 0.023)
                (10000, 0.065)
                (100000, 0.654)
                (1000000, 6.112)
            };
        \addlegendentry{CryptoMask-WO}
        
        \end{axis}
        \end{tikzpicture}
    }
    \subfigure[64-D Representation]{
        \begin{tikzpicture}[scale=0.5]
        \begin{axis}[%
        ylabel=Number,
        scaled ticks=false, 
        legend pos= north west, 
        grid=major,
        grid style=dashed,
        xmin= 1,
        xmax= 10^6,
        ymin= 0,
        ymax= 100000,
        xlabel={Database Size},
        xmode=log,
        log basis x={10},
        ymode=log,
        log basis y={10},
        ylabel={Running Time (s)},
        ]

        \addplot[color=green,mark=square] coordinates {
                (1, 0.032586)
                (10, 0.232492)
                (100, 2.0295)
                (1000, 20.4282)
                (10000, 201.7280)
                (100000, 2030.2791)
                (1000000, 20457.6109)
            };
        \addlegendentry{SFM}       

        \addplot[color=blue,mark=square] coordinates {
                (1, 0.443079)
                (10, 0.454981)
                (100, 0.507175)
                (1000, 0.452591)
                (10000, 1.373210) 
                (100000, 11.529101)
                (1000000, 116.183421)
            };
        \addlegendentry{HERS}

        \addplot[color=red,mark=square] coordinates {
                (1, 0.012)
                (10, 0.02)
                (100, 0.025)
                (1000, 0.094)
                (10000, 0.636)
                (100000, 6.412)
                (1000000, 85.019)
            };
        \addlegendentry{CryptoMask-W}

        \addplot[color=brandeisblue,mark=square] coordinates {
                (1, 0.007)
                (10, 0.007)
                (100, 0.009)
                (1000, 0.024)
                (10000, 0.115)
                (100000, 0.962)
                (1000000, 13.27)
            };
        \addlegendentry{CryptoMask-WO}

        \end{axis}
        \end{tikzpicture}
    }
    \subfigure[128-D Representation]{
        \begin{tikzpicture}[scale=0.5]
        \begin{axis}[%
        ylabel=Number,
        scaled ticks=false, 
        legend pos= north west, 
        grid=major,
        grid style=dashed,
        xmin= 1,
        xmax= 10^6,
        ymin= 0,
        ymax= 100000,
        xlabel={Database Size},
        xmode=log,
        log basis x={10},
        ymode=log,
        log basis y={10},
        ylabel={Running Time (s)},
        ]

        \addplot[color=green,mark=square] coordinates {
                (1, 0.022096)
                (10, 0.204216)
                (100, 2.04609)
                (1000, 20.5224)
                (10000, 210.4875)
                (100000, 2087.3901)
                (1000000, 21023.7327)
            };
        \addlegendentry{SFM}

        \addplot[color=blue,mark=square] coordinates {

                (1, 0.883113)
                (10, 0.896448)
                (100, 0.885061)
                (1000, 0.896120)
                (10000, 2.655189) 
                (100000, 22.569065)
                (1000000, 216.374683)
            };
        \addlegendentry{HERS}

        \addplot[color=red,mark=square] coordinates {
                (1, 0.011)
                (10, 0.018)
                (100, 0.03)
                (1000, 0.111)
                (10000, 0.792)
                (100000, 7.405)
                (1000000, 99.232)
            };
        \addlegendentry{CryptoMask-W}

        \addplot[color=brandeisblue,mark=square] coordinates {
                (1, 0.008)
                (10, 0.007)
                (100, 0.014)
                (1000, 0.042)
                (10000, 0.21)
                (100000, 1.988)
                (1000000, 24.642)
            };
        \addlegendentry{CryptoMask-WO}
        
        \end{axis}
        \end{tikzpicture}
    }
    \subfigure[512-D Representation]{
        \begin{tikzpicture}[scale=0.5]
        \begin{axis}[%
        ylabel=Number,
        scaled ticks=false, 
        legend pos= north west, 
        grid=major,
        grid style=dashed,
        xmin= 1,
        xmax= 10^6,
        ymin= 0,
        ymax= 100000,
        xlabel={Database Size},
        xmode=log,
        log basis x={10},
        ymode=log,
        log basis y={10},
        ylabel={Running Time (s)},
        ]

        \addplot[color=green,mark=square] coordinates {
                (1, 0.023235)
                (10, 0.207317)
                (100, 2.09674)
                (1000, 20.4023)
                (10000, 208.2849)
                (100000, 2045.2498)
                (1000000, 20172.9335)
            };
        \addlegendentry{SFM}

        \addplot[color=blue,mark=square] coordinates {
                (1, 3.51321)
                (10, 3.5385)
                (100, 3.51464)
                (1000, 3.50187)
                (10000, 10.546892) 
                (100000, 87.546875)
                (1000000, 860.736451)
            };
        \addlegendentry{HERS}

        \addplot[color=red,mark=square] coordinates {
                (1, 0.018)
                (10, 0.027)
                (100, 0.052)
                (1000, 0.212)
                (10000, 1.314)
                (100000, 12.124)
                (1000000, 161.444)
            };
        \addlegendentry{CryptoMask-W}

        \addplot[color=brandeisblue,mark=square] coordinates {
                (1, 0.007)
                (10, 0.011)
                (100, 0.03)
                (1000, 0.1)
                (10000, 0.634)
                (100000, 5.962)
                (1000000, 76.641)
            };
        \addlegendentry{CryptoMask-WO}

        \end{axis}
        \end{tikzpicture}
    }
    \caption{Performance of evaluation process.}
    \label{Fig::Total Computation}
\end{figure*}
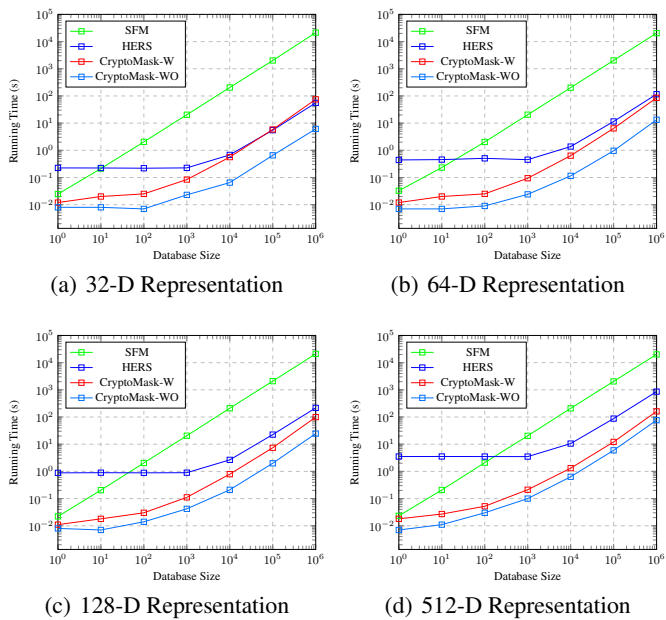

\begin{figure*}[tb]
\centering
    \subfigure[32-D Representation]{
        \begin{tikzpicture}[scale=0.5]
        \begin{axis}[%
        ylabel=Number,
        scaled ticks=false, 
        legend pos= north west, 
        grid=major,
        grid style=dashed,
        xmin= 1,
        xmax= 10^6,
        ymin= 0,
        ymax= 100000,
        xlabel={Database Size},
        xmode=log,
        log basis x={10},
        ymode=log,
        log basis y={10},
        ylabel={Communication Cost (MB)},
        ]

        \addplot[color=green,mark=square] coordinates {
                (1, 0.0870)
                (10, 0.8697)
                (100, 8.6968)
                (1000, 86.9684)
                (10000, 869.6842)
                (100000, 8696.8421)
                (1000000, 86968.4218)
            };
        \addlegendentry{SFM}
        
        \addplot[color=blue,mark=square] coordinates {
                (1, 2.7830)
                (10, 2.7830)
                (100, 2.7830)
                (1000, 2.7830)
                (10000, 8.3490) 
                (100000, 69.5748)
                (1000000, 681.8325)
            };
        \addlegendentry{HERS}

        \addplot[color=red,mark=square] coordinates {
                (1, 0.0874472)
                (10, 0.0878601)
                (100, 0.0917873)
                (1000, 0.27453)
                (10000, 1.19838)
                (100000, 7.07992)
                (1000000, 56.1621)
            };
        \addlegendentry{CryptoMask-W}

        \end{axis}
        \end{tikzpicture}
    }
    \subfigure[64-D Representation]{
        \begin{tikzpicture}[scale=0.5]
        \begin{axis}[%
        ylabel=Number,
        scaled ticks=false, 
        legend pos= north west, 
        grid=major,
        grid style=dashed,
        xmin= 1,
        xmax= 10^6,
        ymin= 0,
        ymax= 100000,
        xlabel={Database Size},
        xmode=log,
        log basis x={10},
        ymode=log,
        log basis y={10},
        ylabel={Communication Cost (MB)},
        ]  
        
        \addplot[color=green,mark=square] coordinates {
                (1, 0.0870)
                (10, 0.8697)
                (100, 8.6968)
                (1000, 86.9684)
                (10000, 869.6842)
                (100000, 8696.8421)
                (1000000, 86968.4218)
            };
        \addlegendentry{SFM}
        
        \addplot[color=blue,mark=square] coordinates {
                (1, 5.566)
                (10, 5.566)
                (100, 5.566)
                (1000, 5.566)
                (10000, 16.698) 
                (100000, 139.15)
                (1000000, 1363.67)
            };
        \addlegendentry{HERS}

        \addplot[color=red,mark=square] coordinates {
                (1, 0.0874414)
                (10, 0.0878363)
                (100, 0.119489)
                (1000, 0.393742)
                (10000, 1.51677)
                (100000, 7.72964)
                (1000000, 58.0634)
            };
        \addlegendentry{CryptoMask-W}
        
        \end{axis}
        \end{tikzpicture}
    }
    \subfigure[128-D Representation]{
        \begin{tikzpicture}[scale=0.5]
        \begin{axis}[%
        ylabel=Number,
        scaled ticks=false, 
        legend pos= north west, 
        grid=major,
        grid style=dashed,
        xmin= 1,
        xmax= 10^6,
        ymin= 0,
        ymax= 100000,
        xlabel={Database Size},
        xmode=log,
        log basis x={10},
        ymode=log,
        log basis y={10},
        ylabel={Communication Cost (MB)},
        ]

        \addplot[color=green,mark=square] coordinates {
                (1, 0.0870)
                (10, 0.8697)
                (100, 8.6968)
                (1000, 86.9684)
                (10000, 869.6842)
                (100000, 8696.8421)
                (1000000, 86968.4218)
            };
        \addlegendentry{SFM}

        \addplot[color=blue,mark=square] coordinates {
                (1, 11.132)
                (10, 11.132)
                (100, 11.132)
                (1000, 11.132)
                (10000, 33.396) 
                (100000, 278.3)
                (1000000, 2727.34)
            };
        \addlegendentry{HERS}

       \addplot[color=red,mark=square] coordinates {
                (1, 0.0874462)
                (10, 0.087842)
                (100, 0.179052)
                (1000, 0.504149)
                (10000, 1.95316)
                (100000, 9.63648)
                (1000000, 61.884)
            };
        \addlegendentry{CryptoMask-W}
        
        \end{axis}
        \end{tikzpicture}
    }
    \subfigure[512-D Representation]{
        \begin{tikzpicture}[scale=0.5]
        \begin{axis}[%
        ylabel=Number,
        scaled ticks=false, 
        legend pos= north west, 
        grid=major,
        grid style=dashed,
        xmin= 1,
        xmax= 10^6,
        ymin= 0,
        ymax= 100000,
        xlabel={Database Size},
        xmode=log,
        log basis x={10},
        ymode=log,
        log basis y={10},
        ylabel={Communication Cost (MB)},
        ]
        
        \addplot[color=green,mark=square] coordinates {
                (1, 0.0870)
                (10, 0.8697)
                (100, 8.6968)
                (1000, 86.9684)
                (10000, 869.6842)
                (100000, 8696.8421)
                (1000000, 86968.4218)
            };
        \addlegendentry{SFM}

        \addplot[color=blue,mark=square] coordinates {
                (1, 44.528)
                (10, 44.528)
                (100, 44.528)
                (1000, 44.528)
                (10000, 133.584) 
                (100000, 1113.2)
                (1000000, 10909.36)
            };
        \addlegendentry{HERS}

        \addplot[color=red,mark=square] coordinates {
                (1, 0.08743)
                (10, 0.115446)
                (100, 0.321364)
                (1000, 0.998883)
                (10000, 3.46144)
                (100000, 13.994)
                (1000000, 76.157)
            };
        \addlegendentry{CryptoMask-W}
        \end{axis}
        \end{tikzpicture}
    }
    \caption{Communication overhead comparison of our protocol with SFM and HERS.}
    \label{Fig::Total Communication}
\end{figure*}
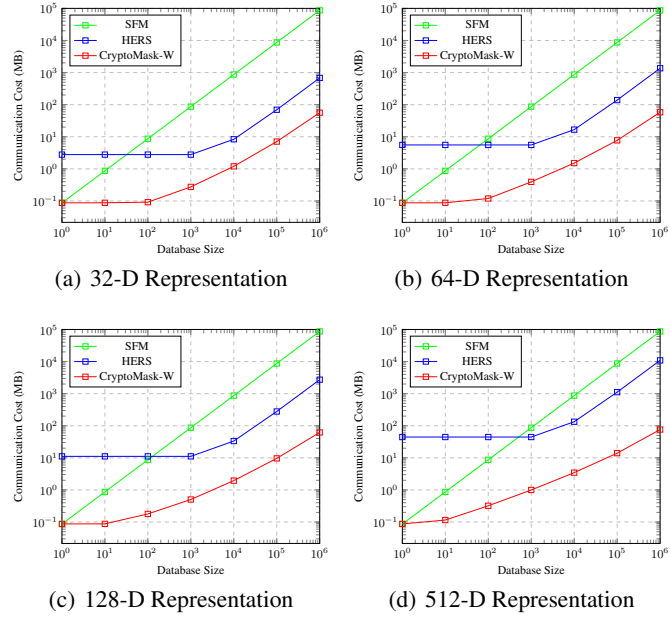

\subsection{Efficiency}
Following the same dataset construction from~\cite{engelsma2022hers}, we evaluate \sys~on four representations at different dimensions (32-D, 64-D, 128-D, and 512-D).
Fig.~\ref{Fig::Total Computation} and Fig.~\ref{Fig::Total Communication} separately report the concrete computation and communication overhead with dataset sizes varying from 1 to 100 million. In the following, for simplicity, we use SFM to name the work in~\cite{boddeti2018secure} and use HERS to name the work in~\cite{engelsma2022hers}.

\noindent \textbf{Computation overhead.}
We report two computation overhead lines of CryptoMask in Fig.~\ref{Fig::Total Computation} where CryptoMask-W denotes we fully implement CryptoMask while CryptoMask-WO represents the version without the secure result-revealing protocol. 
In particular, CryptoMask-WO, SFM, and HERS have comparable information leakage, where they all leak the computed similarity to the verifier. 

From Fig.~\ref{Fig::Total Computation} we can see both CryptoMask-W and CryptoMask-WO outperform SFM in the four dimensions settings. 
The reason is the primary computation overhead in secure face recognition is caused by the homomorphic multiplication, which is $m$ times in~\cite{boddeti2018secure} while it is $\lceil \frac{m}{N-d}\rceil d $ times in CryptoMask. 
Compared with HERS, CryptoMask-WO shows the same tendency but enjoys less computation overhead. The main reason is we provide optimizations for computation. 
As for CryptoMask-W, the required computation overhead is near to HERS but achieves better security by concealing the similarity between face vectors from the verifier. 
CryptoMask is sensitive to the feature dimension, and the running time gap between SFM and CryptoMask-W drops with the increase of the dimension. For example, when working on 32-D, CryptoMask-W outperforms SFM by $283 \times$ against a gallery of 100 million. When working on 512-D, CryptoMask-W only saves around $132 \times$ computation than SFM, yet CryptoMask still shows its high efficiency for the large-scale dataset. Even when compared with similar work HERS, CryptoMask-W lies between CryptoMask-WO and HERS, indicating that it enjoys a better computation overhead while ensuring database security. 

\noindent \textbf{Communication.}
Fig.~\ref{Fig::Total Communication} details the communication consumption of CryptoMask-W, SFM and HERS. It shows that CryptoMask-W requires the least communication resource than the other two. The main reason comes from the given communication optimizations mentioned in Section~\ref{subsec::optimization}. 

\section{Conclusion} \label{sec:conclusion}

We introduce CryptoMask, a practical privacy-preserving face recognition protocol that leverages homomorphic encryption and secure multi-party computation techniques. Our encoding strategy facilitates an efficient enrollment process, enabling DP to add more face vectors to CS. We construct an efficient matrix computation for distance calculation, based on our encoding method. Unlike existing state-of-the-art techniques that reveal the computed distance to the verifier, we protect intermediate results using a secure result-revealing protocol. Our experiments show that CryptoMask outperforms existing approaches in both computation and communication.


\section*{Acknowledgment}
We thank the anonymous reviewers for their insightful comments and suggestions. 
Bai and Russello would like to acknowledge the MBIE-funded programme STRATUS (UOWX1503) for its support and inspiration for this research.

%
%
%
\bibliographystyle{splncs04}
\bibliography{ref}

\appendix
\section{Complexity and Security Analysis}\label{sec:security}
We first provide a theoretical complexity analysis to show the efficiency of \sys. 
Then we show that \sys~is secure against a semi-honest adversary while assuming KG is fully trusted.

\subsection{Complexity Analysis}
In \sys, communication overhead mainly comes from two parts. One is from CS, who sends all the encrypted distances to the verifier, which contains $O(Nm/d)$ communication cost. Another one is the result of the secure revealing process, which requires $O(ml)$ communication. We can obtain the overall communication complexity as $O(Nm/d+ml)$. The computation overhead is more complex. We set the computation for data encryption using HE as $C_{en}$, for homomorphic multiplication as $C_{mul}$, for homomorphic addition as $C_{add}$, for key switching as $C_{sw}$, for secure comparison as $C_{com}$ and for secure~\textbf{B2A} as $C_{cov}$. The overall computation overhead for the CS side is $O((Nm/d)(C_{com}+C_{add} + C_{sw}) + m(C_{com}+C_{cov}))$ and for the verifier side is $O(C_{en} + m(C_{com}+C_{cov}))$.

\subsection{Security Analysis}


\textbf{Privacy of Face Vector Matrix}. 
In \sys, all face vectors are encrypted by HE, and only the KG knows the secret key. Due to the semantic security of HE, neither CS nor the verifier learns sensitive information about the underlying encrypted face vector; thus, the privacy of the face vector is always maintained. 

Now we show \sys~only reveals a face recognition result to the verifier and nothing else to either party. 
This is argued as regards to a corrupted CS and a corrupted verifier, respectively. Note we only provide the security of the HE-based part as the simulation of the comparison/\textbf{B2A} protocols can be implemented in the existing ways.

\textbf{Corrupted CS.} We first demonstrate the security against a semi-honest CS. Intuitively, the security against a semi-honest CS comes from the fact that the CS's view of the execution includes only ciphertext, thus reducing the argument to the semantic security of HE. We now give the formal argument. 

Let $\mathcal{A}$ be the semi-honest CS in the real protocol. We construct a simulator $\mathcal{S}$ in the ideal world as follows:

\begin{enumerate}
    \item[1.]At the beginning of the protocol execution, $\mathcal{S}$ receives the input $\boldsymbol{\mathsf{A}}$ from the environment $\mathcal{E}$ and also receives the public key $pk$ and the vector length $d$. The simulator sends $\boldsymbol{\mathsf{A}}$ to the trusted party.
    \item[2.]Start running $\mathcal{A}$ on input $\boldsymbol{\mathsf{A}}$. Next, $\mathcal{S}$ computes and sends a ciphertext $ct$, which encrypts a $d$ dimensional vector $\boldsymbol{0}$ to the CS under the public key $pk$.
    \item[3.]Output whatever $\mathcal{A}$ outputs.
\end{enumerate}
We argue the above simulated view is indistinguishable from real protocol execution. 
Using the fact that $\mathcal{A}$ is semi-honest, at the end of the protocol in the real world, the verifier obtains the encryption of $\boldsymbol{\mathsf{A}} \cdot \boldsymbol{b}$ where $\boldsymbol{b}$ is the verifier's queried face image. Since $\mathcal{S}$ is semi-honest, this also holds in the ideal world. Since $\boldsymbol{\mathsf{A}} \cdot \boldsymbol{b}$ is a deterministic function, the joint distribution of the verifier's output and the adversary's output decomposes. Thus, it is sufficient to show that the simulated view from $\mathcal{S}$ is computationally indistinguishable from the real view from $\mathcal{A}$.

The view of $\mathcal{A}$ in the real world contains one part: the encrypted face image ${ct}$ from the verifier. When interacting with the simulator $\mathcal{S}$, adversary $\mathcal{A}$ sees an encryption of $\boldsymbol{0}$. Security follows immediately by the semantic security of the BFV scheme.

\textbf{Corrupted Verifier.} We now prove the security against a semi-honest verifier. We construct a simulator $\mathcal{S}$ in the ideal world as follows:

\begin{enumerate}
    \item[1.]At the beginning of the execution, $\mathcal{S}$ receives the input $\boldsymbol{b}$ from the environment $\mathcal{E}$ and also receives the BFV key pairs $(pk,sk)$ and the matrix size $m,d$. The simulator sends $\boldsymbol{b}$ to the trusted party.
    \item[2.]Start running $\mathcal{A}$ on input $\boldsymbol{b}$. Next, $\mathcal{S}$ computes and sends ciphertexts $c_i$ which is the encryption of an $m \times d$ matrix filled by some random values to the verifier under the public key $pk_v$.
    \item[3.]Output whatever $\mathcal{A}$ outputs.
\end{enumerate}
At the end of face recognition, CS has no output. Thus, to show the security against a semi-honest verifier, it suffices to show that the output of $\mathcal{S}$ is computationally indistinguishable from the output of the adversary $\mathcal{A}$. Now we show the view of simulator $\mathcal{S}$ in the ideal world is computationally indistinguishable from the view of the adversary $\mathcal{A}$ in the real world.

The view of $\mathcal{A}$ in the real world contains one part: the encrypted face database $\{c_1, \cdots, c_n\}$ from CS. When interacting with the simulator $\mathcal{S}$, adversary $\mathcal{A}$ sees the encryption of random values. Security follows immediately by the semantic security of the BFV scheme.

\section{Accuracy} \label{appendix:accuracy}
We report the results of face recognition on dataset LFW for state-of-the-art face representation FaceNet in Table~\ref{Table::Accuracy}. We only test face templates of 128-D. For more results on different representations, we refer to~\cite{boddeti2018secure}, which is also constructed on BFV. Same as~\cite{boddeti2018secure}, we report true acceptance rate~(TAR) at three different operating points of $0.01\%, 0.1\%$ and $1.0\%$ false accept rates~(FARs). 
We first report the performance of the unencrypted face images. 
We treat these outputs as a baseline to compare. To evaluate encrypted face images, we consider four different quantization for each element in facial features. Specifically, we employ precision of 0.1, 0.01, 0.0025 and 0.0001. It shows that the performance of most given precision is competitive with the performance conducted from the raw data. We conclude that CryptoMask working over HE and MPC can perform as well as the one working over raw data.

\begin{table}[]
\footnotesize
\centering
\caption{F{\upshape ace recognition accuracy for LWF dataset} (TAR @ FAR in $\%$)}\label{Table::Accuracy}
\begin{tabular}{c|ccc}
\hline
\multirow{2}{*}{\textbf{Method}} & \multicolumn{3}{c|}{128-D FaceNet (Accuracy)}                    \\ \cline{2-4} 
                        & \multicolumn{1}{c|}{0.01\%} & \multicolumn{1}{c|}{0.1\%} & 1\%   \\ \hline
No FHE                  & \multicolumn{1}{c|}{98.70}  & \multicolumn{1}{c|}{98.70} & 98.70 \\ \hline
FHE($1.0\times10^{-4}$)          & \multicolumn{1}{c|}{98.70}  & \multicolumn{1}{c|}{98.70} & 98.70 \\ \hline
FHE($2.5\times10^{-3}$)         & \multicolumn{1}{c|}{98.70}  & \multicolumn{1}{c|}{98.70} & 98.70 \\ \hline
FHE($1.0\times10^{-2}$)          & \multicolumn{1}{c|}{98.76}  & \multicolumn{1}{c|}{98.76} & 98.76 \\ \hline
FHE($1.0\times10^{-1}$)          & \multicolumn{1}{c|}{98.50}  & \multicolumn{1}{c|}{98.50} & 98.50 \\ \hline
\end{tabular}
\end{table}

\end{document}